\documentclass[prb,floatfix,showpacs,preprint,preprintnumbers]{revtex4}
\usepackage{amsmath,amssymb}
\usepackage{graphicx,color}
\usepackage{longtable}

\begin{document}
\title{Electronic and Vibrational Properties of $\gamma$-AlH$_3$ }
\author{Yan Wang}
\author{Jia-An Yan}
\author{M. Y. Chou}
\affiliation{School of Physics, Georgia Institute of Technology, Atlanta,
             Georgia 30332-0430}
\date{\today}
\begin{abstract}
Aluminum hydride (alane) AlH$_3$ is an important material in
hydrogen storage applications. It is known that AlH$_3$ exists in
multiply forms of polymorphs, where  $\alpha$-AlH$_3$ is found to be
the most stable with a hexagonal structure. Recent experimental
studies on $\gamma$-AlH$_3$ reported an orthorhombic structure with
a unique double-bridge bond between certain Al and H atoms. This was
not found in $\alpha$-AlH$_3$ or other polymorphs. Using density
functional theory, we have investigated the energetics, and the
structural, electronic, and phonon vibrational properties for the
newly reported $\gamma$-AlH$_3$ structure. The current calculation
concludes that $\gamma$-AlH$_3$ is less stable than $\alpha$-AlH$_3$
by 2.1 KJ/mol. Interesting binding features associated with the
unique geometry of $\gamma$-AlH3 are discussed from the calculated
electronic properties and phonon vibrational modes.  The binding of
H-s with higher energy Al-$p,d$ orbitals is enhanced within the
double-bridge arrangement, giving rise to a higher electronic energy
for the system. Distinguishable new features in the vibrational
spectrum of $\gamma$-AlH$_3$ were attributed to the double-bridge
and hexagonal-ring structures.
\end{abstract}
 \maketitle

\section{Introduction}

Aluminum hydride AlH$_3$ (alane) is among the most promising metal
hydrides for the hydrogen storage medium, containing a usable
hydrogen fraction of 10.1 wt.$\%$ with a density of 1.48 g/ml. The
enthalpy of the reaction is found to be low, resulting in minimal
heat exchange for both charging and discharging reactions. This
material is thermodynamically unstable near ambient conditions, but
it is kinetically stable without releasing much hydrogen for years.
However, extremely high hydrogen pressure (exceeding 25 kbar) is
required to achieve charging.  The decomposition of AlH$_3$ occurs
in a single step:
\begin{equation}
 AlH_3 \rightarrow Al + \frac{3}{2}H_2.
\end{equation}
How to overcome the kinetic barriers and to find new routes to
synthesize AlH$_3$ for reversibility have been the focus of many
research activities recently.

Early studies identified seven polymorphs of AlH$_3$: $\alpha$,
$\alpha'$, $\beta$, $\delta$, $\varepsilon$, $\gamma$, and
$\xi$.\cite{brower76} It was suggested experimentally that the
$\alpha$ phase is the most stable, followed by the $\beta$ and
$\gamma$ phases.\cite{reilly1,reilly2}  The decomposition of the
$\gamma$ and $\beta$ phases is faster than that of $\alpha$ phase.
An reaction enthalpy of 7.1 KJ/mol-H$_2$ and 11.4 KJ/mol-H$_2$ for
$\gamma$ and $\alpha$, respectively, is
reported.\cite{reilly1,reilly2} The exothermic transition to the
$\alpha$ phase from the $\gamma$ phase is also reported.  Only the
$\alpha$ polymorph AlH$_3$ has been experimentally investigated
extensively, including structural characterization,\cite{alpha}
thermodynamic measurements,\cite{therm1,therm2} and thermal and
photolytic kinetic
studies.\cite{therm2,kinet1,kinet2,kinet3,kinet4,kinet5,kinet6}
Theoretical calculations\cite{cal1,cal2,cal3} of $\alpha$-AlH$_3$
from first principles have been performed to study the structural
stability and electronic and thermodynamic properties.  Limited
studies have been conducted on other polymorphs and their
properties.

Recently, the crystal structure of $\gamma$-AlH$_3$ is reported by
two separate groups using synchrotron X-ray powder
diffraction\cite{gamma} and powder neutron diffraction,\cite{gamma2}
respectively. A unique feature of double-bridge bonds involving
Al-2H-Al is identified in addition to the normal bridge bond of
Al-H-Al as found in $\alpha$-AlH$_3$.  In the present study, using
density functional theory and the linear response approach, we
investigate the electronic properties and phonon spectra for the
newly published $\gamma$-AlH$_3$ crystal
structure.\cite{gamma,gamma2} The phase stability and interesting
binding characteristics in $\gamma$-AlH$_3$ are presented and
compared with that of $\alpha$-AlH$_3$. The origin of the cohesive
energy difference in these two phases is discussed. Distinct phonon
vibrational modes arising from the double-bridge and hexagonal ring
structures are identified.

\section{Calculational procedures }

The calculations are based on density functional theory.\cite{dft}
The Kohn-Sham equations are solved in a plane-wave basis  using the
Vienna {\it ab initial} simulation package (VASP).\cite{met1,met2}
For the exchange-correlation functional, the generalized gradient
approximation (GGA) of Perdew and Wang (PW91)\cite{met3} is
employed. The electron-ion interaction is described by ultrasoft
pseudopotentials (USPP).\cite{metus} The $k$ space integrals are
evaluated using the  sampling generated by the Monkhorst-Pack
procedure.\cite{kpoint}  The calculation for both $\alpha$- and
$\gamma$-AlH$_3$ structures are performed with a $k$-point mesh of
$7\times 7 \times 7$. The relaxations of cell geometry and atomic
positions are carried out using a conjugate gradient algorithm until
the Hellman-Feynman force on each of the unconstrained atoms is less
than 0.01 eV/$\AA$. The nuclear coordinates are first allowed to
relax while the cell volume is fixed at the experimental value. Then
simultaneous relaxations of the cell volume, shape and atom
coordinates are conducted. For the total energy calculations, the
plane-wave energy cutoff is 600 eV. The self-consistent total energy
converges to within $10^{-5}$ eV/cell.

The vibrational properties are studied with density functional
perturbation theory within the linear response.\cite{linear} The
dynamical matrices are obtained for a uniform grid of $q$ vectors of
$4\times 4 \times 4$ and $3\times 3\times 3$ over the Brillouin zone
of $\alpha$- and $\gamma$-AlH$_3$, respectively. This dynamical
matrix is then Fourier-transformed to real space and the
force-constant matrices are constructed, which are used to obtain
the phonon frequencies. The PWSCF numerical code\cite{pwscf} was
used in our calculations for the zero-point energies and phonon
density of states.

\section{Results and Discussions }
\subsection{structure and Energetics}

\begin{figure}[h]
\centering
\includegraphics[clip=true,width=5cm]{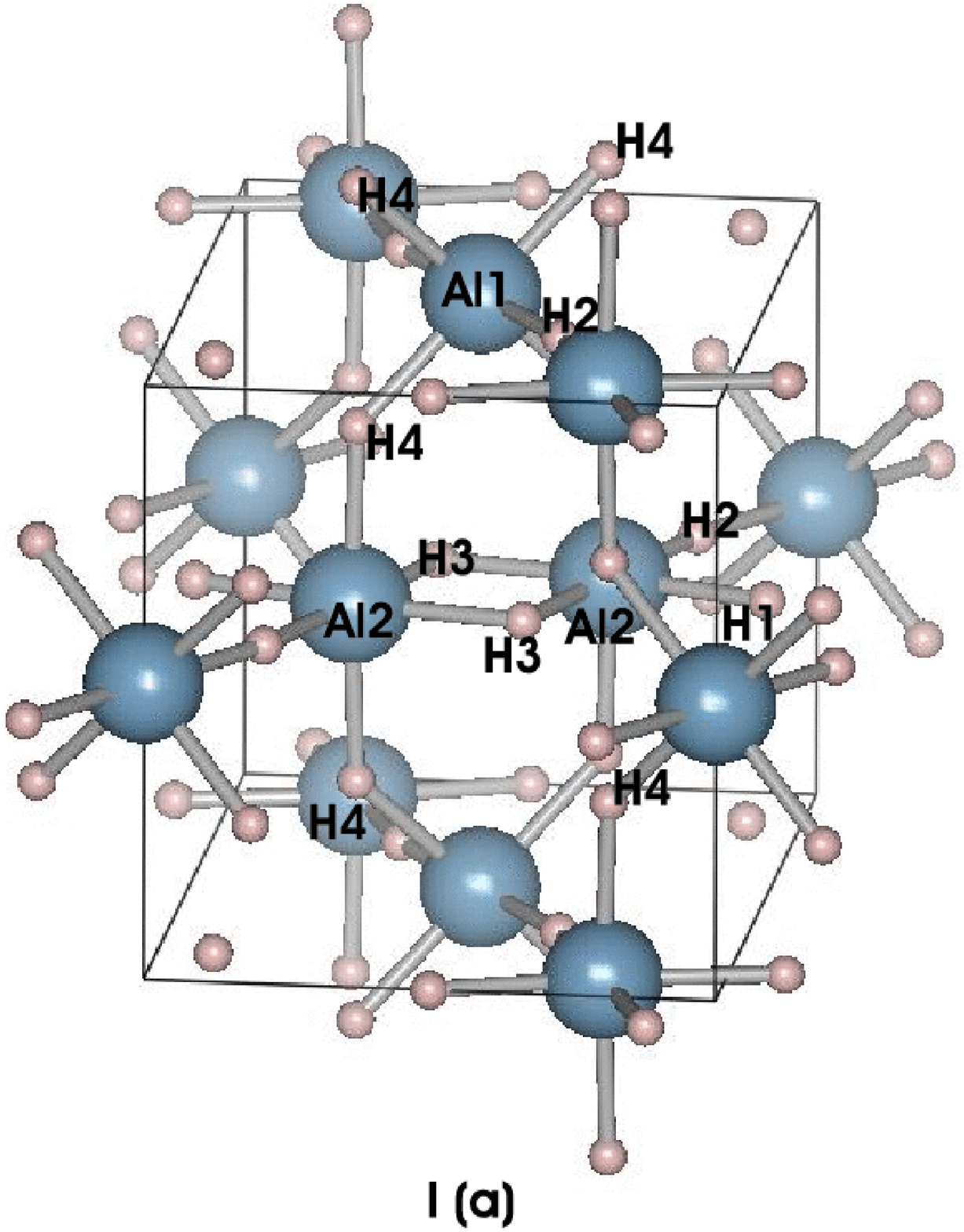}
\includegraphics[clip=true,width=5cm]{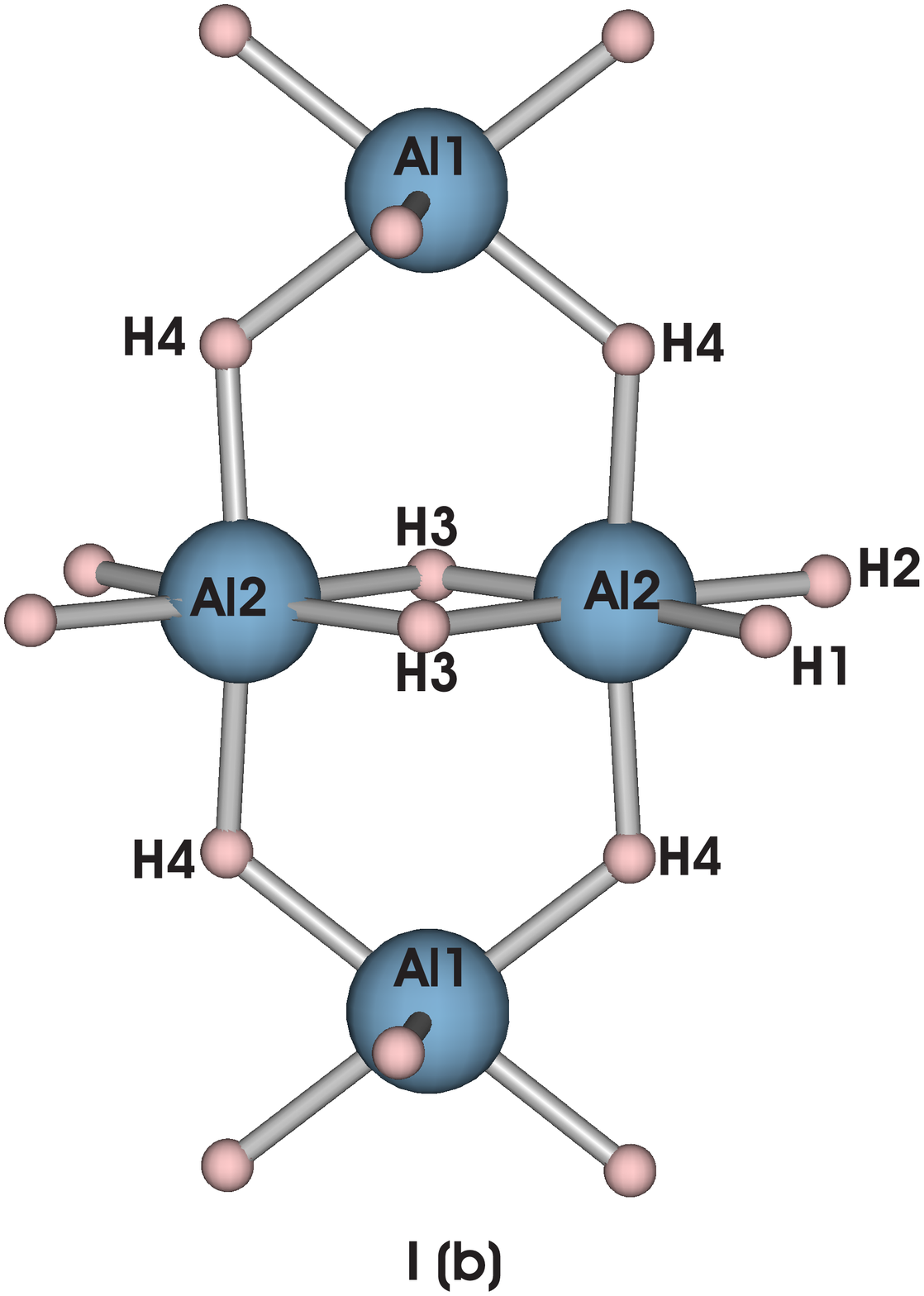}
\includegraphics[clip=true,width=5cm]{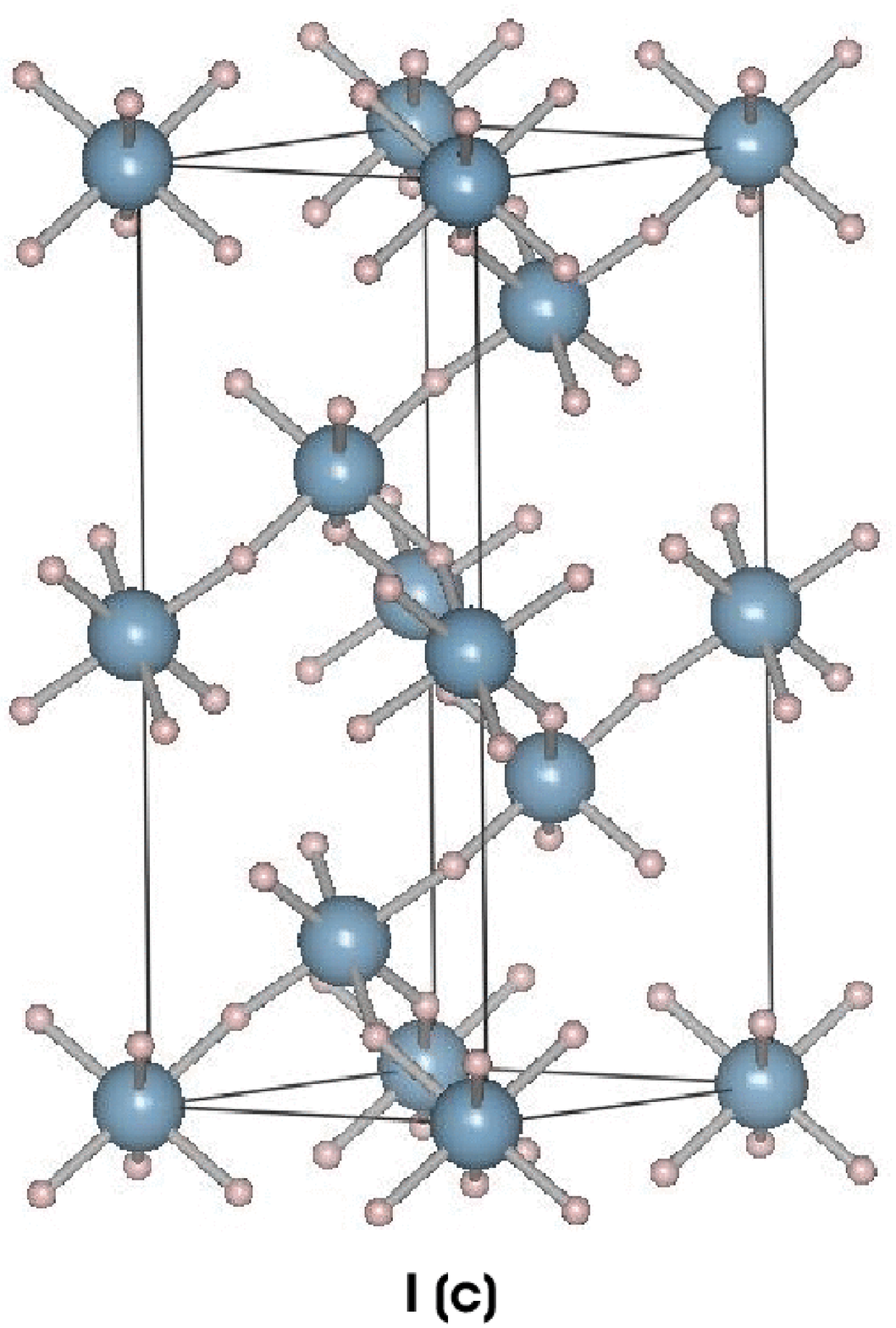}
\caption{(Color Online) Crystal structures of (a) orthorhombic
$\gamma$-AlH$_3$, (b) hexagonal-ring structure in $\gamma$-AlH$_3$,
and (c) hexagonal $\alpha$-AlH$_3$. The large and small spheres
denote Al and H atoms, respectively. } \label{fig:struct}
\end{figure}

The structural characterization from an earlier high-resolution
synchrotron X-ray diffraction\cite{alpha} study concluded that
$\alpha$-AlH$_3$ has a rhombohedral lattice of space group
$R\bar{3}c$ (No. 167).  Recent diffraction
experiments\cite{gamma,gamma2} determined that $\gamma$-AlH$_3$ has
an orthorhombic unit cell with space group $Pnnm$ (No. 58). The unit
cell for $\gamma$-AlH$_3$ is shown in Figure \ref{fig:struct}(a),
compared with the hexagonal $\alpha$-AlH$_3$ in Figure
\ref{fig:struct}(c).  The experimentally reported lattice constants
are listed in Table I.  The building element for both phases is the
AlH$_6$ octahedron, where one Al atom is surrounded by six H atoms.
However, the packing scheme in $\gamma$-AlH$_3$ is more complex than
that in $\alpha$-AlH$_3$. The AlH$_6$ octahedra are connected simply
by sharing vertices in $\alpha$-AlH$_3$, as illustrated in Figure
\ref{fig:struct}(c).  The network of these octahedra produces only
one type of Al-H-Al bridge bond which has a bond angle and a bond
length of 142$^\circ$ and 1.712 ${\AA}$, respectively. In contrast,
in $\gamma$-AlH3 the AlH$_6$ octahedra are connected not only by
sharing vertices as in $\alpha$-AlH$_3$, but also by sharing edges,
as shown in Figure \ref{fig:struct}(a). As a consequence, two
nonequivalent Al atoms, Al1 and Al2, are created. Four nonequivalent
H atoms (H1-H4)\cite{gamma} are also identified as shown in Figure
\ref{fig:struct}(a), while only one type H atom exists in
$\alpha$-AlH$_3$.  Al1 is involved with the normal bridge bond with
H, which is similar to that in $\alpha$-AlH$_3$ but with a slight
difference in the bond length and angle.  However, Al2 involves a
new type of double-bridge configuration Al2-2H3-Al2, as shown near
the center of unit cell in Figure \ref{fig:struct}(a). This
double-bridge bond gives a smaller distance of $2.60$ ${\AA}$
between the two Al2 atom compared with the Al-Al separation of 3.24
${\AA}$ in $\alpha$-AlH$_3$ and 2.86 ${\AA}$ in Al metal.  In
addition to the new double-bridge configuration,  a hexagonal-ring
structure is found consisting of two Al2, two Al1, and four H4
atoms, as shown in Figure \ref{fig:struct}(b). The four Al atoms are
on one plane parallel to the c-axis while the H4 atoms have a slight
displacement ($\sim$0.14 $\AA$) out of this plane. These rings are
connected to form linear chains along c-axis. The structures of
double bridge in the ab-plane and the hexagonal-ring along c-axis
are unique to $\gamma$-AlH$_3$ and have not been found in any other
hydrogen-containing aluminum compounds. Furthermore, the $\alpha$
phase possesses a higher order of symmetry than that of the $\gamma$
phase, with a smaller primitive trigonal unit cell. The total number
of formula units (f.u.) in a unit cell is six and two for the
$\gamma$ and $\alpha$ phases, respectively. The molecular volume
(the density) for the $\gamma$ phase is found to be higher (lower)
than that of the $\alpha$ phase by 11$\%$ (10$\%$).

\begin{table}[h]
\caption{Calculated cohesive energies and structural parameters for
$\gamma$-AlH$_3$ and $\alpha$-AlH$_3$ compared with values from
experiments and previous calculations. }\centering
\begin{ruledtabular}
\begin{tabular}{ccccccccc}
Configuration&&Cohesive Energy&\multicolumn{6}{c}{Lattice Parameters}\\
&&eV/(f.u.)&a(\AA)&b(\AA)&c(\AA)&$\alpha(^\circ)$&$\beta(^\circ)$&$\gamma(^\circ)$\\
\hline
$\alpha$-AlH$_3$:&&&&&&&& \\
Current work&&-14.052&4.49&4.49&11.80&90.0&90.0&120.0\\
Previous work&(Ref.\onlinecite{cal1})&&4.42&4.42&11.80&90.0&90.0&120.0\\
&(Ref.\onlinecite{cal2})&&4.489&4.4489&11.820&90.0&90.0&120.0\\
Experiment&(Ref.\onlinecite{alpha})&&4.449&4.449&11.813&90.0&90.0&120.0\\
$\gamma$-AlH$_3$:&&&&&&&& \\
Current work&&-14.028&5.43&7.40&5.79&90.0&90.0&90.0\\
Experiment&(Ref.\onlinecite{gamma})&&5.3806&7.3555&5.77509&90.00&90.00&90.00\\
&(Ref.\onlinecite{gamma2})&&5.3672&7.3360&5.7562&90.00&90.00&90.00\\

\end{tabular}
\end{ruledtabular}
\label{eorder}
\end{table}

\begin{table}[h]
\caption{  Interatomic distances ($\AA$) and bond angles (deg)
obtained from the fully relaxed structure of $\gamma$-AlH$_3$. The
values from the synchrotron X-ray diffraction\cite{gamma} and the
powder neutron diffraction\cite{gamma2} are also included for
comparisons. }\centering
\begin{ruledtabular}
\begin{tabular}{ccccccccc}
Bond&Calc.&Ref.16/Ref.17&Bond&Calc.&Ref.16/Ref.17&Angle&Calc.&Ref.16/Ref.17\\
\hline
Al1-Al2&3.18&3.1679/3.155&Al2-H1&1.69&1.668/1.657&Al1-H4-Al2&134.7&124.0/133.88\\
Al2-Al2&2.62&2.602/2.585&Al2-H2&1.70&1.664/1.678&Al1-H2-Al2&168.8&171.0/169.99\\
Al1-H2&1.76&1.769/1.764&Al2-H3&1.72&1.70/1.755&Al2-H1-Al2&180.0&180.0/179.99\\
Al1-H4&1.72&1.784/1.696&Al2-H4&1.72&1.790/1.733&Al2-H3-Al2&97.3&100.7/97.53\\
&&&H3-H3&2.30&2.16/2.266&H4-Al1-H4&85.9&76.34/85.00\\
&&&&&&H4-Al2-H4&176.3&170.63/176.84\\
\end{tabular}
\end{ruledtabular}
\label{gdistance}
\end{table}

Using the experimentally established space groups and unit cells as
shown in Figure \ref{fig:struct}, the total energies and structural
parameters for both the $\alpha$ and $\gamma$ phases are studied
from first-principles calculations. The calculated results for the
fully relaxed structures are summarized in Table I. The theoretical
lattice parameters are in good agreement with the observed values
for both $\gamma$- and $\alpha$-AlH$_3$ and are consistent with the
results from previous $\alpha$-AlH$_3$ calculations.\cite{cal1,cal2}
The calculated cohesive energies indicate that $\gamma$-AlH$_3$ is
energetically less stable than $\alpha$-AlH$_3$ by 2.3 KJ/mol, which
is in fair agreement with the measured value of 4.3
KJ/mol.\cite{reilly1,reilly2} In addition, in $\gamma$-AlH$_3$
various bond lengths for non-equivalent Al and H atoms are evaluated
and compared with recent experimental results of synchrotron X-ray
powder diffraction\cite{gamma} and powder neutron
diffraction\cite{gamma2} in Table II. Good agreements are found
between the calculated and observed values. Without including the
zero point energy, the calculated reaction enthalpies for equation
(1) of $\alpha$ and $\gamma$ phases are 9.2 and 10.8 KJ/mol-H$_2$,
respectively, which are also consistent with the measured
values.\cite{reilly1,reilly2}

\begin{figure}[!hbp]
\centering
\includegraphics[width=3in]{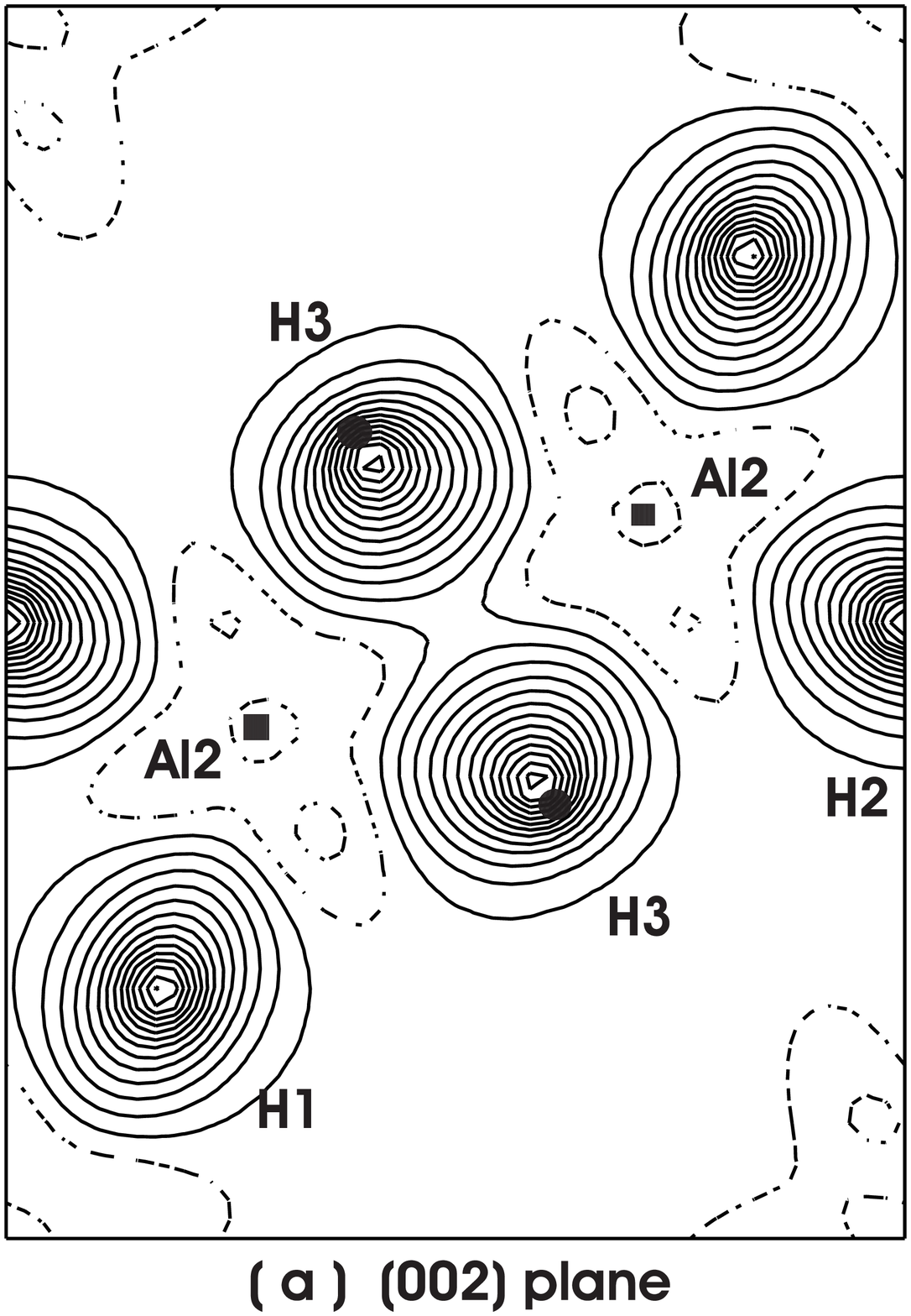}
\includegraphics[width=3in]{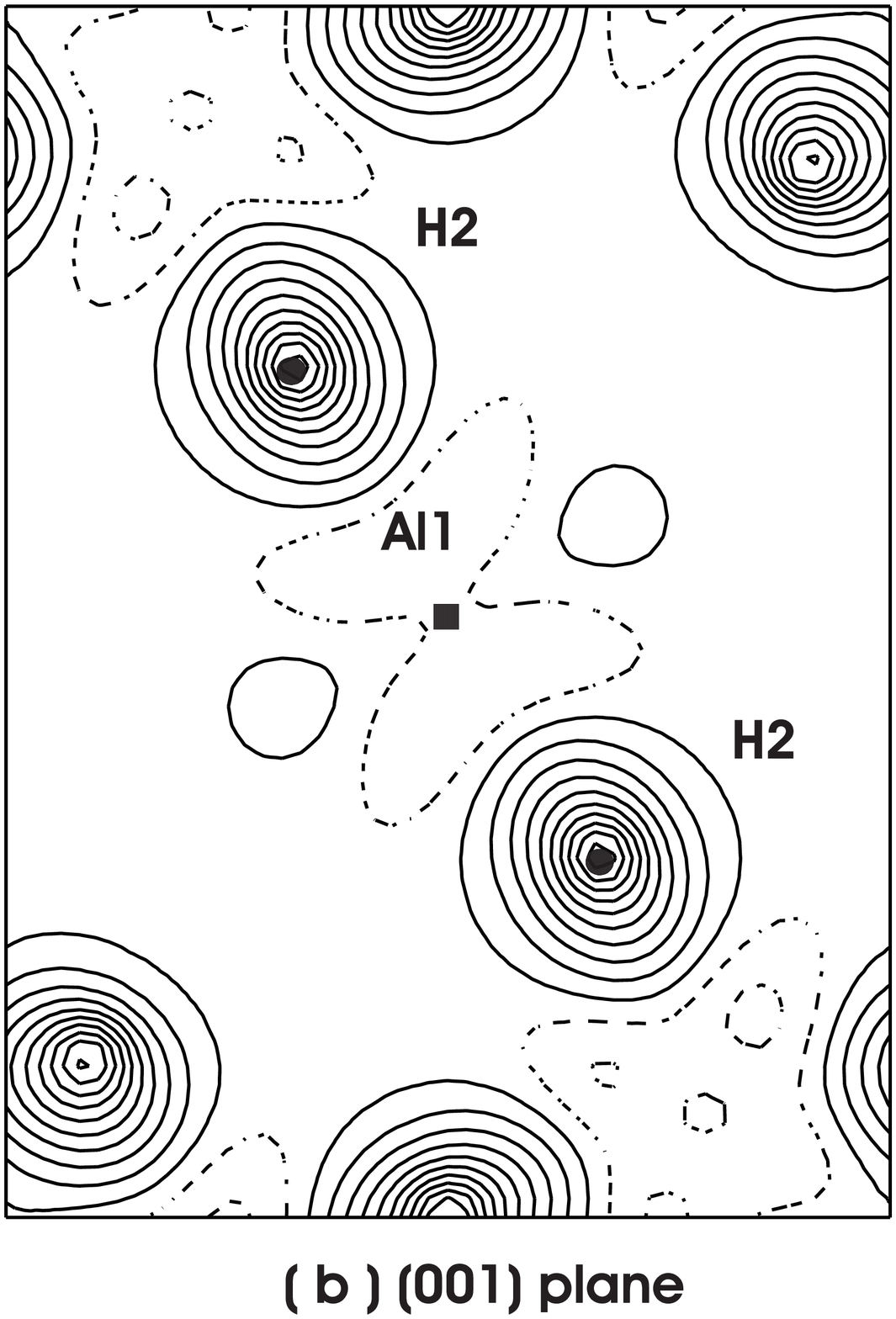}
\caption{Difference charge density plots (see text) for
$\gamma$-AlH$_3$ in the planes containing Al and H atoms: (a) the
(002) plane and (b) the (001) plane. The solid squares and circles
represent Al and H positions, respectively. The density contour
interval is 0.001 electrons/({\AA}$^3$).  Charge deficiency is
represented by dashed lines, while the density increase near the
hydrogen atoms is represented by  solid lines. } \label{fig:gchg}
\end{figure}

\begin{figure}[!hbp]
\centering
\includegraphics[width=5in]{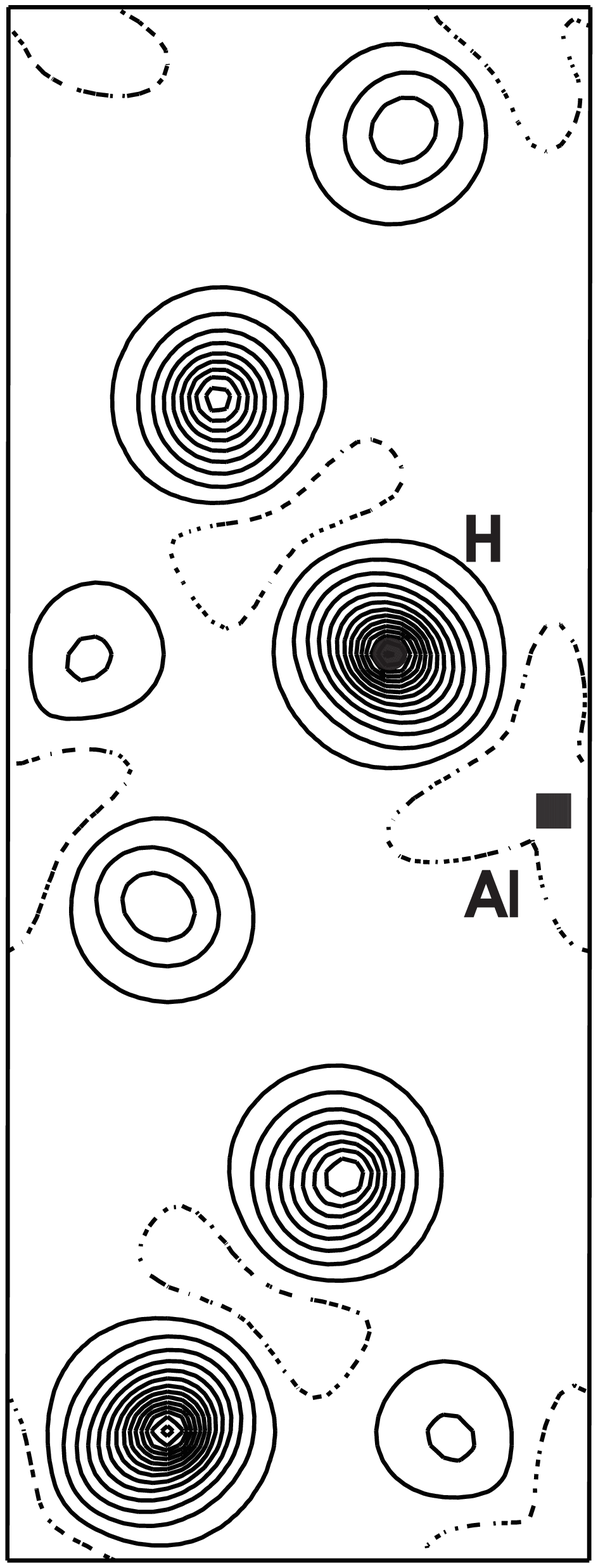}
\caption{Charge-density difference plot for $\alpha$-AlH$_3$ in a
plane containing Al and H atoms. The solid square and circle
represent Al and H positions, respectively. The density contours in
an interval of 0.001 electrons/({\AA}$^3$) are presented for the
(010) plane. Charge deficiency is represented by dashed lines, while
the density increase near the hydrogen atoms is represented by solid
lines.} \label{fig:achg}
\end{figure}

\subsection{Electronic Structure}

The electronic structure of the $\gamma$ - AlH$_3$ is first analyzed
by examining the charge distribution and charge transfer.  The
difference charge density $\Delta \rho({\bf r})$ is the difference
between the total charge density of the solid and a superposition of
the atomic charge with the same spatial coordinates as in the solid:
\begin{equation}
 \Delta \rho({\bf r})=\rho_{solid}({\bf r}) - \sum_{i} \rho_{atom}^{i}({\bf r - R_i}),
\end{equation}
where the sum is over all the atoms. Figure \ref{fig:gchg} shows
such plots in the planes containing two different types of Al-H
configurations in $\gamma$-AlH$_3$. The solid squares and circles
represent Al and H positions, respectively. The (002) plane shown in
Figure \ref{fig:gchg}(a) contains one double-bridge bond Al2-2H3-Al2
and two other H atoms (H1 and H2), while  the (001) plane shown in
Figure \ref{fig:gchg}(b) contains a normal bridge H2-Al1-H2. The
difference density plot shows the positive values (solid contours)
at the H positions, indicating a charge transfer from Al to H. As a
result, aluminum is positively and hydrogen is negatively charged.
The maximum of contour lines is in the order of 0.014
electrons/({\AA})$^3$ for both (a) and (b). The minimum is about
-0.002 electrons/({\AA})$^3$ and the step size of the contours is
0.001 electrons/({\AA})$^3$. The zero difference charge density line
forms a closed contour around H, leaving  a negative charge density
in the interstitial regions.

In Figure \ref{fig:gchg}(a), note that the positive contours of
difference charge density are not exactly centered at the two H3
sites.  The local maximum near each H3 is slightly shifted toward
each other and the zero difference charge density line encloses both
H3, showing some interactions between the H3 pair. In contrast, this
is not seen for other H atoms, nor in the normal H-Al-H structure in
$\alpha$-AlH$_3$. The separation between two H3 atoms in a
double-bridge configuration is 2.3 {\AA}, which is slightly smaller
than that of other neighboring H pairs. The distance of the two Al2
atoms (2.62 {\AA}) is also small compared with that in Al metal
(2.86 {\AA}). The contour lines in Figure \ref{fig:gchg}(a) also
shows enhanced interactions between H-$s$ and Al-$d$ electrons. For
comparison, the difference charge density for $\alpha$-AlH$_3$ is
also calculated and plotted in Figure \ref{fig:achg} for the (010)
plane containing normal Al-H bonds. It is found that the normal
bridge bond in $\gamma$-AlH$_3$ [Figure \ref{fig:gchg}(b)] has
similar characteristics as that in $\alpha$-AlH$_3$ (Figure
\ref{fig:achg}). The binding between Al and H atoms involves a
charge transfer in both $\alpha-$ and $\gamma$-AlH$_3$.

\begin{figure}[!hbp]
\centering
\includegraphics[width=5in]{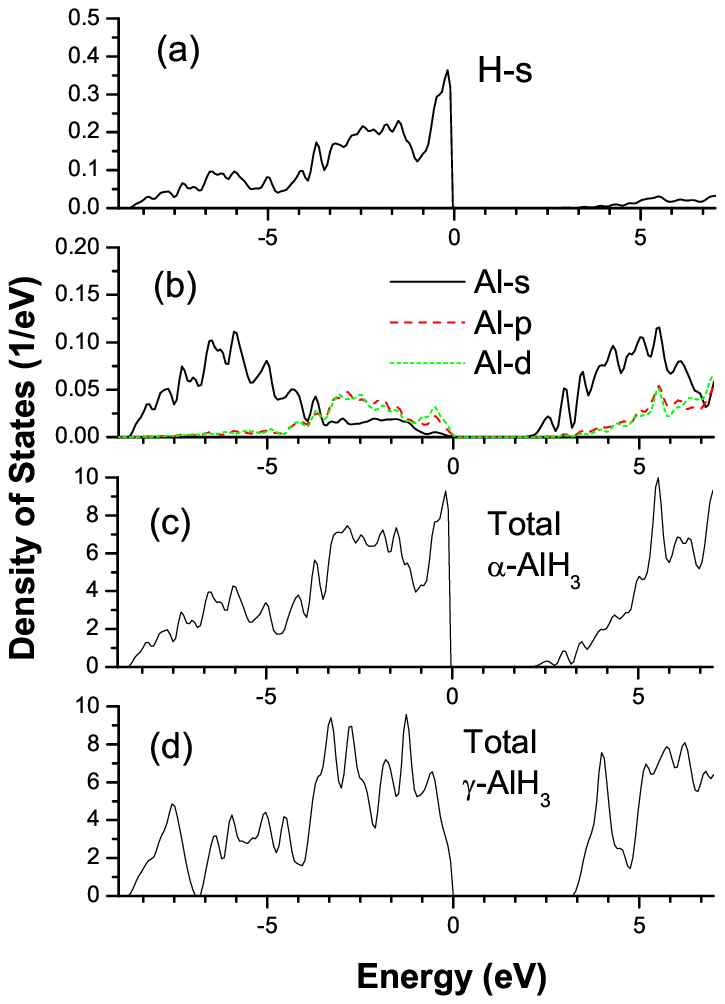}
\caption{(Color Online) Calculated electronic density of states
(DOS) projected onto (a) H and (b) Al in $\alpha$-AlH$_3$, as well
as the total DOS for the (c) $\alpha$-AlH$_3$ and (d)
$\gamma$-AlH$_3$. The Fermi level is at zero.} \label{fig:apdos}
\end{figure}

\begin{figure}[!hbp]
\centering
\includegraphics[width=5in]{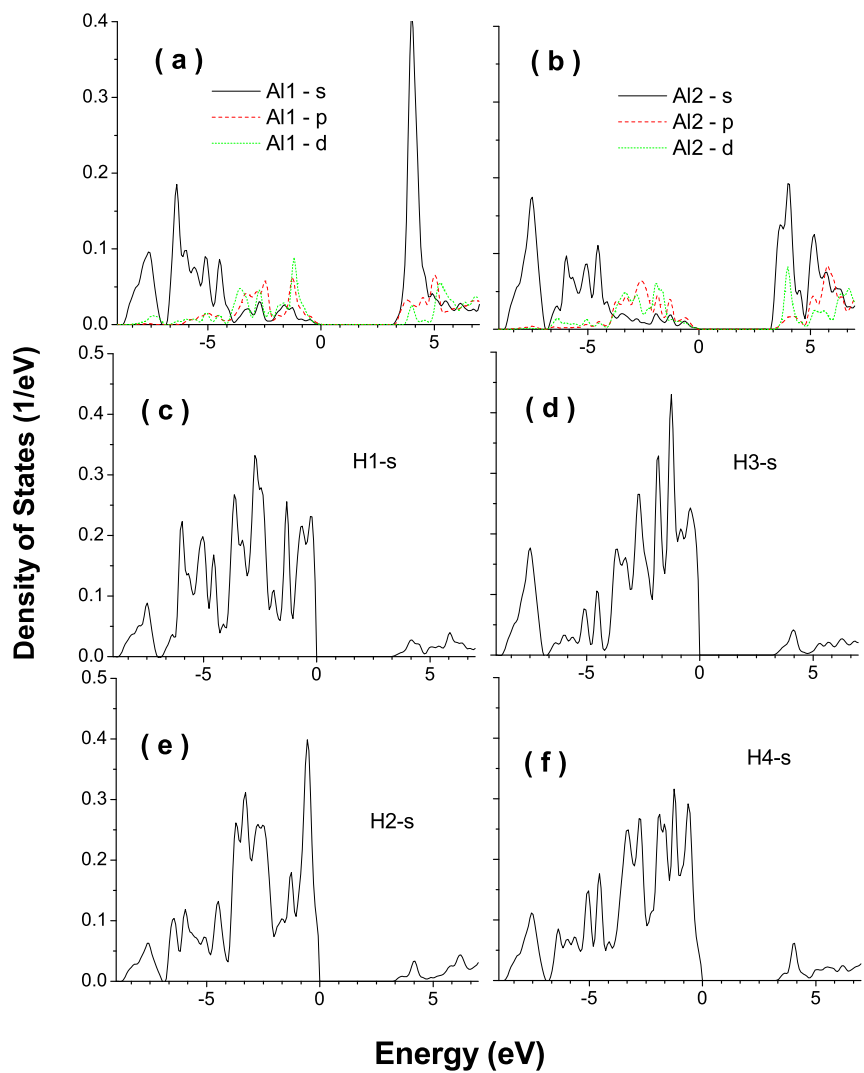}
\caption{(Color Online) Calculated electronic density of states
projected onto non-equivalent atoms in $\gamma$-AlH$_3$. The Fermi
energy is set to be zero.} \label{fig:gpdos}
\end{figure}

\begin{figure}[!hbp]
\centering
\includegraphics[width=5in]{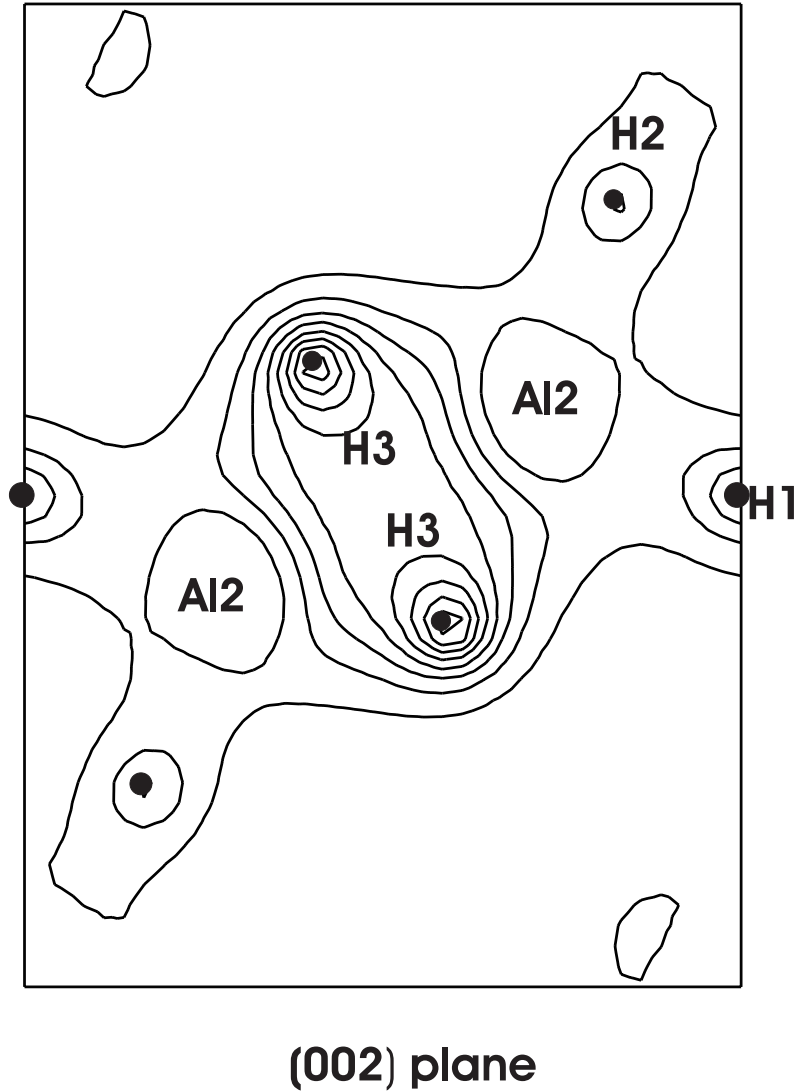}
\caption{Total charge density for the lowest two energy bands
projected onto the (002) plane, the same plane as shown in Figure
\ref{fig:gchg}(a). The solid circles represent H positions. The
maximum, minimum  and the interval of the contours are 0.85, 0.002,
and 0.1 electrons/({\AA}$^3$),respectively.} \label{fig:pchg2d}
\end{figure}

\begin{figure}[!hbp]
\centering
\includegraphics[width=6in]{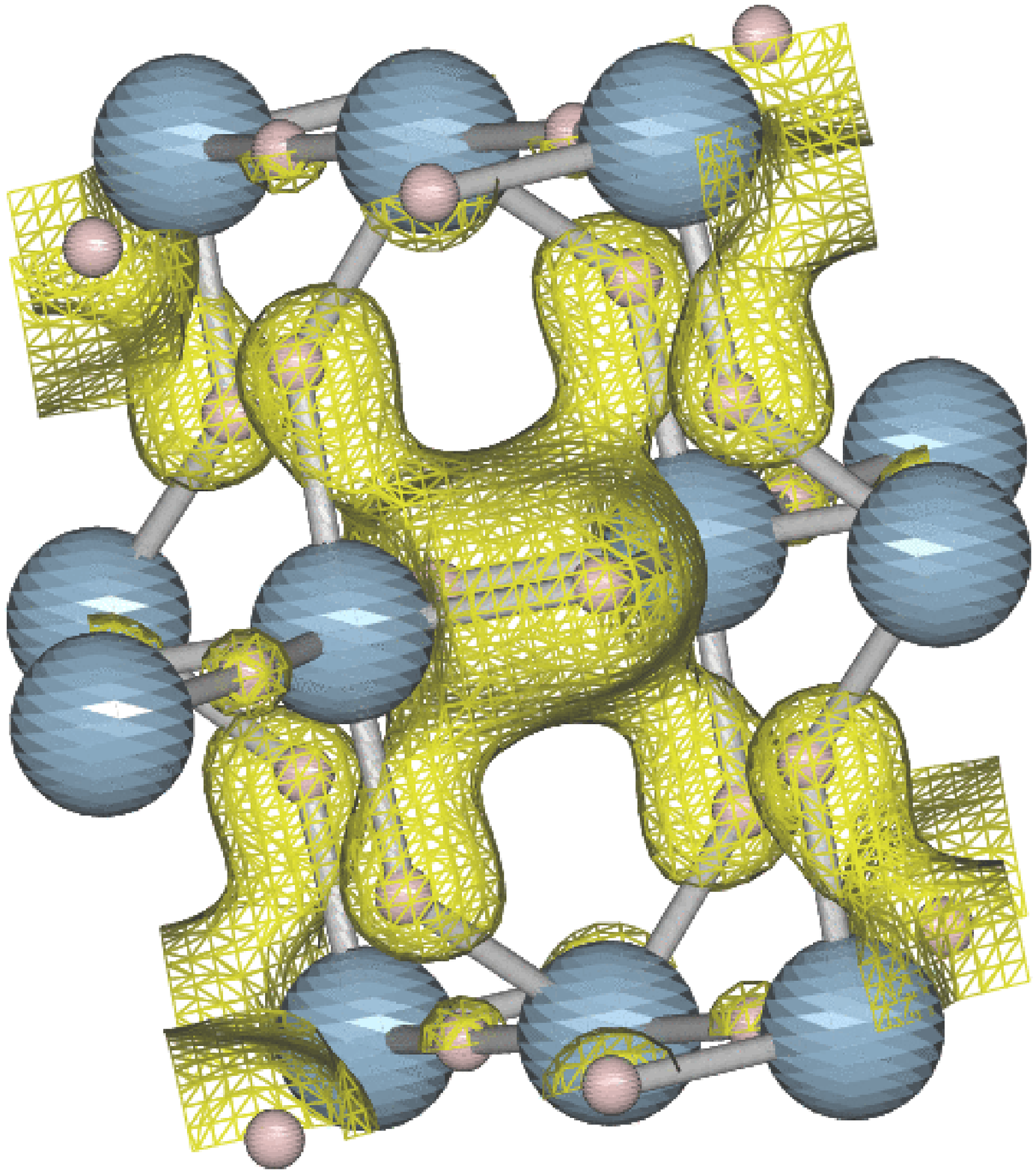}
\caption{(Color Online) Total charge density isosurface for the
lowest two energy bands in $\gamma$-AlH$_3$. The cell structure is
the same as shown in Figure \ref{fig:struct}(a). The constant charge
density surface shown is for 0.45 electrons/$(\AA)^3$. The charge
accumulation within the hexagonal-ring structure defined in Figure
\ref{fig:struct}(b) is clearly shown.} \label{fig:pchg}
\end{figure}

The total electronic densities of states (DOS) for both structures
are calculated and given in Figure \ref{fig:apdos}. The projected
DOS onto H and Al  for $\alpha$-AlH$_3$ are also presented in Figure
\ref{fig:apdos}. The angular-momentum projected DOS is evaluated by
integrating over a sphere centered at each atom with a radius of 1.0
{\AA} for Al and 0.9 {\AA} for H, respectively. The choice of the H
radius is based on the charge distribution around H in order to
catch the charge transfer to H. The radius of Al is then chosen
according to the Al-H bond length in order to cover as much of the
interstitial region as possible without overlapping the spheres. The
Fermi energy is set to zero in these plots.

The current results for $\alpha$-AlH$_3$  are comparable with
previous calculations.\cite{cal1,cal2,cal3} The H-s component spans
the whole energy range. The lowest peak from $\sim$ -8.7 to -4.0 eV
is found to correspond to  H-s and Al-s states, while the
higher-energy features are composed of H-s and Al-p or -d states.
The total bond width is similar in both $\alpha$- and
$\gamma$-AlH$_3$. A difference between the two is clearly seen near
-7 eV, where a small band gap is created in the low energy region in
$\gamma$-AlH$_3$. This new gap gives rise to a broad, separated DOS
peak from -8.7 to -7.0 eV.

In order to understand the new feature found in the DOS of
$\gamma$-AlH$_3$, the projected density of states onto
non-equivalent Al and H atoms are analyzed in Figure
\ref{fig:gpdos}. The same radius values used  in $\alpha$-AlH$_3$
are adopted for integrating over the spheres around Al and H atoms.
The valence states below the Fermi energy can be grouped into three
energy regions: the low( -8.7 to -7.0), the middle(-6.8 to -4.0),
and the high( -4.0 to 0 eV) energy regions.

The broad peak in the low energy region of -8.7 to -7.0 is composed
of considerable H3-s states, with some s contributions from other H
and two Al atoms. This broad peak consists of the first two energy
bands separated from the other bands at higher energies by a small
gap. In Figure \ref{fig:pchg2d}, the total charge density for these
two bands is calculated and plotted for the (002) plane, the same
plane shown in Figure \ref{fig:gchg}(a). The maximum, minimum and
interval of the contours are 0.85, 0.002, and 0.1
electrons/$(\AA)^3$, respectively. It shows noticeable charge
accumulation around the H3-H3 pair. The three-dimensional isosurface
of the total charge density for the same lowest two bands is shown
in Figure \ref{fig:pchg}, which has a density value of 0.45
electrons/$(\AA)^3$. Again, the interaction between the H3-H3 pair
is illustrated. In addition, considerable charge is found around
four H4 atoms within the same hexagonal-ring as defined in Figure
\ref{fig:struct}(b). Therefore, the broad peak in the low energy
region is attributed mainly to the unique double-bridge and
hexagonal-ring structures in $\gamma$-AlH$_3$.

In fact, the projected DOS associated with H3 is reduced in the
middle energy region (-6.8 to -4.0 eV) compared with other H atoms.
The missing amplitude is shifted to both the low (-8.7 to -7.0 eV)
and high (-4.0 to 0 eV) energy regions. The former contains an
increased H3-H3 interaction, the H-s and Al2-s coupling, while the
latter exhibits an enhanced interaction between H3-s and Al2-p and
-d states. The stronger interaction of H3-s states with
higher-energy Al2-p and -d states due to geometry gives rises to a
higher electronic energy for the system.

\subsection{Vibrational Properties}

It is expected that the high vibrational frequencies of light
hydrogen atoms may result in a significant zero-point energy, which
needs to be included in studying the ground-state properties. In
order to consider the zero-point energy $E_{zpt}$ contributions to
the energetics between the two AlH$_3$ phases, the phonon density of
states (DOS) are evaluated for both $\alpha$- and $\gamma$-AlH$_3$.
Although the absolute value of E$_{zpt}$ is noticeable for the two
phases, the difference turns out to be small $\Delta E_{zpt} =
E_{zpt}(\alpha) - E_{zpt}(\gamma) = 0.2$ KJ/mol. By including this
correction to the total energies previously listed in Table I, the
$\gamma$ phase has a slightly higher energy than the $\alpha$ phase
by 2.1 KJ/mol.

\begin{figure}[!p]
\centering
\includegraphics[width=5in]{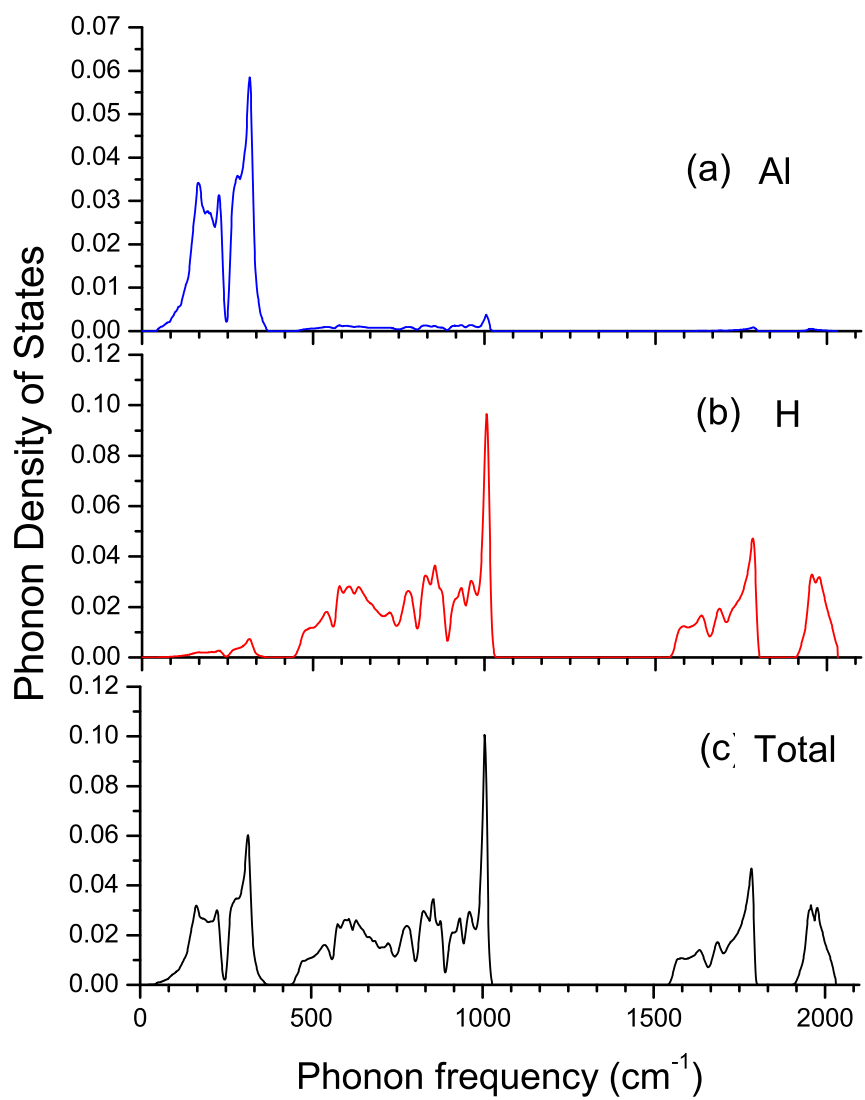}
\caption{(Color Online) Total and partial phonon density of states
(DOS) for $\alpha$-AlH$_3$.} \label{fig:alphonon}
\end{figure}

\begin{figure}[!hb]
\centering
\includegraphics[width=6in]{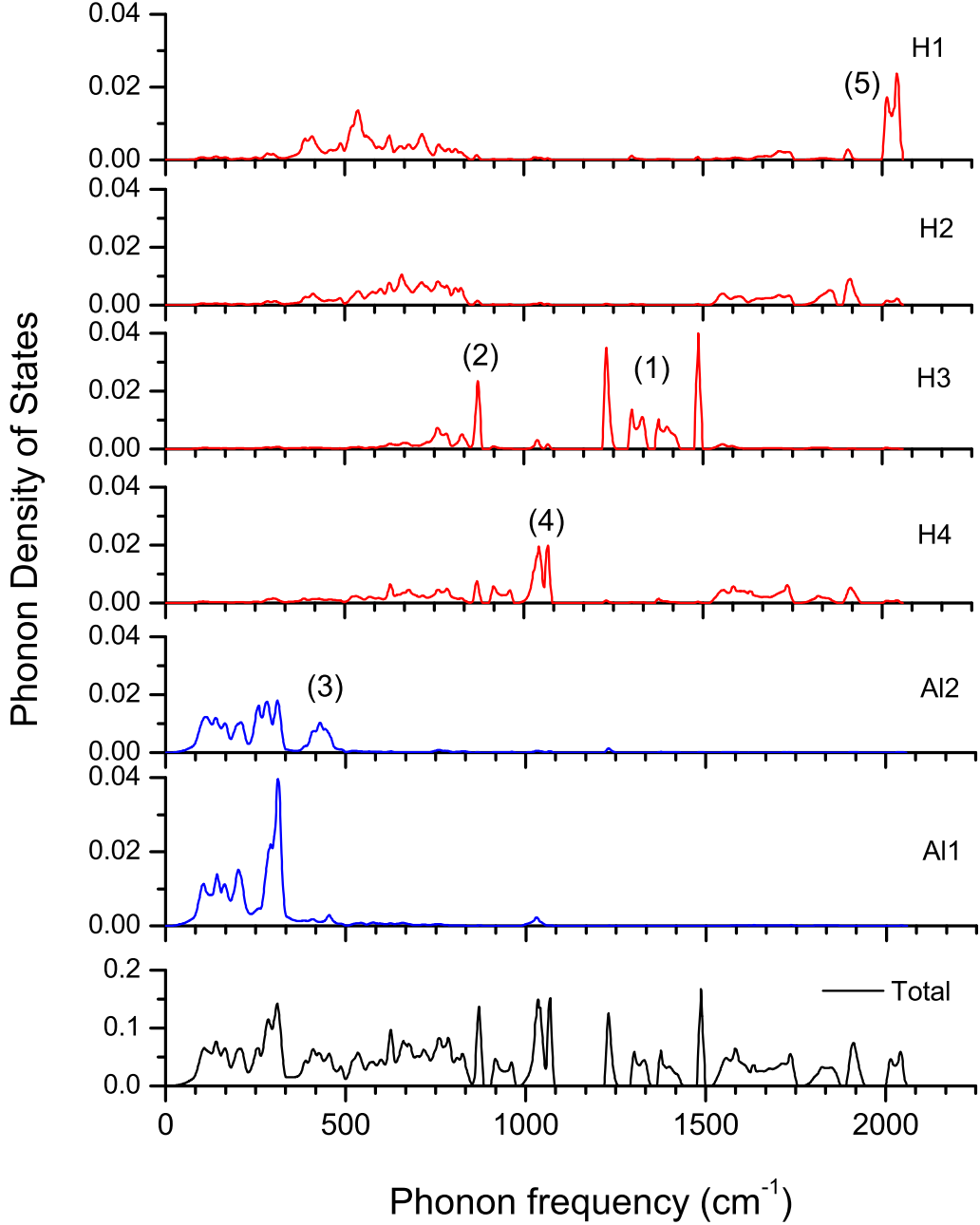}
\caption{(Color Online) Total and partial phonon density of states
(DOS) for $\gamma$-AlH$_3$.} \label{fig:gphonon}
\end{figure}

The partial phonon density of states (DOS) for atom $\tau$ is
defined as:
\begin{equation}
\rho_{\tau}(\omega)=\sum_{q}\sum_{j=1}^{3N}|{\bf e}_{\tau}({\bf q},
j)|^2 \delta(\omega - \omega({\bf q},j)),
\end{equation}
where the N is the total number of atoms per unit cell, ${\bf q}$ is
the phonon momentum, $j$ labels the phonon branch, ${\bf
e}_{\tau}({\bf q}, j)$ is the phonon polarization vector for atom
$\tau$, and $\omega({\bf q},j)$ is the phonon frequency.  The total
and projected phonon density of states for $\alpha$- and $\gamma$-
AlH$_3$ are presented in the Figures \ref{fig:alphonon} and
\ref{fig:gphonon}, respectively. Our calculated phonon DOS for
$\alpha$-AlH$_3$ is in good agreement with  previous
calculations,\cite{cal1,cal2} showing three distinct groups in the
whole frequency range. The decomposed phonon DOS indicates that the
low frequency modes below 350 cm$^{-1}$ are mainly from Al atoms,
while the middle and high frequency modes are H- dominated. This is
due to the large mass ratio between H atom to Al atom. The
high-frequency phonons above 1550 cm$^{-1}$ are associated with H
motion in the Al-H bond stretching modes.  No vibrational modes was
found in the frequency region from 1025 to 1550 cm$^{-1}$, yielding
gap of 525 cm$^{-1}$ in $\alpha$-AlH$_3$.

The phonon DOS for $\gamma$-AlH$_3$ is more complex than that for
$\alpha$-AlH$_3$ due to the structural difference. The $\gamma$-
phase unit cell has much more atoms (24) than the $\alpha$-phase
(8), therefore the calculated phonon DOS for $\gamma$-AlH$_3$ in
Figure \ref{fig:gphonon} exhibit more features. Comparing with the
phonon DOS of $\alpha$-AlH$_3$, although the modes located in the
three frequency regions remain, multiple new additional peaks appear
in the spectrum.  In order to understand how different atomic
species contribute to these vibrations, the partial phonon DOS is
calculated for the four types of H and two types of Al atoms, as
illustrated in Figure \ref{fig:gphonon}.

The new features labeled in Figure \ref{fig:gphonon} can be
summarized as follows: (1) four new vibrational peaks appear within
the gap region of the $\alpha$-AlH$_3$ spectrum in the frequency
range of 1200 to 1500 cm$^{-1}$. These are vibrational modes of H3
in the plane of the Al2-2H3-Al2 complex. It consists of both bond
stretching and bending motions by H3 on this plane. (2) A new strong
narrow peak is introduced in the middle frequency region around a
frequency of 875 cm$^{-1}$, corresponding to the H3 vibrations
perpendicular to the Al2-2H3-Al2 plane. (3) An additional broaden
peak appears in the low frequency region from 375 to 475 cm$^{-1}$,
which is dominated by Al2 in-plane vibrational modes, coupled with
some contributions from H1 and H2 atoms with little contribution
from H3. The displacements of the two Al2 atoms have the same
magnitude but opposite directions, indicating a paired motion
between them. As a consequence, these modes have higher frequencies
than the other Al modes. The first three new features are largely
related to the motions of the two H3 and the two Al2 atoms in the
double-bridge structure. (4) A set of two peaks around 1550
cm$^{-1}$ are associated with the out-of-plane motion of four H4
atoms in the hexagonal-ring structure shown in Figure
\ref{fig:struct}(b). (5) The set of high-frequency peaks around 2012
cm$^{-1}$ are associated with H1 atoms which connect two Al2 atoms
in opposite directions. The angle Al2-H1-Al2 is 180$^\circ$. The
vibrational modes of H1 are in the $ab$ plane and along the
Al2-H1-Al2 line. These new features displayed in the
$\gamma$-AlH$_3$ phonon DOS can serve as indicators of these special
Al and H arrangements.

\section{Conclusion}

We have performed pseudopotential density-functional calculations to
study the electronic and phonon vibrational properties for the newly
reported aluminum hydride $\gamma$-AlH$_3$ structure. The calculated
structural parameters are in good agreement with results from the
diffraction experiments. Our energetic study for the AlH$_3$ system
indicates that $\gamma$-AlH$_3$ is less stable than $\alpha$-AlH$_3$
by $\sim$ 2 KJ/mol.  The unique double-bridge configuration in
$\gamma$-AlH$_3$ is investigated by examining the electronic
properties and phonon vibrational modes. It was found that the
double-bridge arrangement modified the binding between H and Al
atoms. The projected DOS indicates that interactions exist between
the double-bridge H3 atoms and that the interaction between H3-$s$
and higher-energy Al2-$p$ and -$d$ states is enhanced. The latter
yields a higher electronic energy in $\gamma$-AlH$_3$. New features
in phonon vibrational spectrum associated with double-bridge bonds
and hexagonal-ring complex were also identified.

\section{Acknowledgment}
We thank Dr. V. A. Yartys for bringing Ref. \onlinecite{gamma} to
our attention and stimulating discussions. This work is supported by
the Department of Energy under grant No. DE-FG02-05ER46229.

\end{document}